\documentclass[
aps, 
prb, 
10pt,  
amssymb, 
amsmath, 
superscriptaddress, 
tightenlines, 
twocolumn, 
notitlepage, 
] {revtex4-2}

\usepackage{graphicx}
\usepackage{dcolumn}
\usepackage{bm}
\newcommand{\equalcontrib}{These authors contributed equally.}

\begin{document}

\title{Chiralometer: Direct Torque Detection of Crystal Chirality}

\author{Nikolai Peshcherenko}\thanks{\equalcontrib}
\affiliation{%
Max Planck Institute for Chemical Physics of Solids, 01187, Dresden, Germany}

\author{Ning Mao}\thanks{\equalcontrib}
\affiliation{Max Planck Institute for Chemical Physics of Solids, 01187, Dresden, Germany}

\author{Claudia Felser}
\affiliation{%
Max Planck Institute for Chemical Physics of Solids, 01187, Dresden, Germany}

\author{Yang Zhang}
\affiliation{Department of Physics and Astronomy, University of Tennessee, Knoxville, TN 37996, USA}
\affiliation{Min H. Kao Department of Electrical Engineering and Computer Science, University of Tennessee, Knoxville, Tennessee 37996, USA}


\begin{abstract}
Chirality governs phenomena ranging from chemical reactions to the topology of quasiparticle charge carriers. However, a direct macroscopic probe for crystal chirality remains a significant challenge, especially in time reversal symmetric systems with weak circular dichroism signal. Here, we propose the ``Chiralometer'', a mechanical detection method that probes chirality by driving angular momentum carriers out of equilibrium. Using first-principles calculations and semiclassical transport theory, we demonstrate that a temperature gradient in insulators or an electric field in metals induces uncompensated angular momentum in phonons and electrons, respectively. This imbalance generates a macroscopic mechanical torque ($\tau \sim 10^{-11} N \cdot m$) well within the sensitivity of modern torque magnetometry and cantilever-based sensors.
We identify robust signatures in chiral crystals such as Te, SiO$_2$, and the topological semimetal CoSi. Our work establishes mechanical torque as a fundamental order parameter for chirality, offering a transformative tool for orbitronics and chiral quantum materials.

\end{abstract}

\maketitle

\textbf{Introduction} 
Chirality dictates the fundamental symmetry and topological landscape of electronic excitations in condensed matter. Its decisive role spans a broad spectrum of phenomena, including circular dichroism~\cite{circular_dichroism1,circular_dichroism2,circular_dichroism3,circular_dichroism4,circular_dichroism5,circular_dichroism6}, magnetochiral anisotropy \cite{magnetochiral1,magnetochiral2,magnetochiral3,magnetochiral4} and chirality induced spin selectivity effect \cite{CISS_1,CISS_2,CISS_3,CISS_4,CISS_5,CISS_6,CISS_7}. While these phenomena typically rely on structural chirality, which is defined by the non-superimposability of a system and its mirror image, the concept is equally vital in the context of topological edge states and band crossings~\cite{weyl_crossing1}. Such crossings generate non-trivial Berry curvature and can manifest as the chiral anomaly, often probed via large negative longitudinal magnetoresistance~\cite{NLMR1,NLMR2,NLMR3,NLMR4,NLMR5} in topological semimetals.

However, established experimental signatures often lack clarity or provide only an indirect connection to chiral physics. For instance, the chirality-induced spin selectivity effect remains a rather indirect phenomenon, detected via the inverse spin Hall effect. Beyond that, negative longitudinal magnetoresistance can be induced by node-unrelated contributions such as current jetting~\cite{current_jetting1,current_jetting2,current_jetting3}. Even nonlinear Hall effects, which arise from the Berry curvature dipole in noncentrosymmetric materials, can be sensitive to extrinsic scattering. Consequently, developing a macroscopic, direct experimental probe that establishes chirality as a primary observable is essential.

In this work, we propose a method to directly detect crystal chirality via mechanical torque measurements, a concept we term the ``Chiralometer''. Our approach generalizes the Einstein--de Haas effect of spin angular momentum~\cite{einstein_de_haas,EdH_review} to time-reversal-symmetric systems by probing orbital angular momentum properties in the absence of a magnetic field.
In thermal equilibrium, time-reversal symmetry enforces the exact cancellation of opposite-momentum contributions: $\mathbf{l}(\mathbf{q}) = -\mathbf{l}(-\mathbf{q})$. However, driving a system out of equilibrium induces a nonequilibrium distribution of angular-momentum carriers, such as electrons, magnons, or phonons.

\begin{figure}
    \centering
    \includegraphics[width=0.9\linewidth]{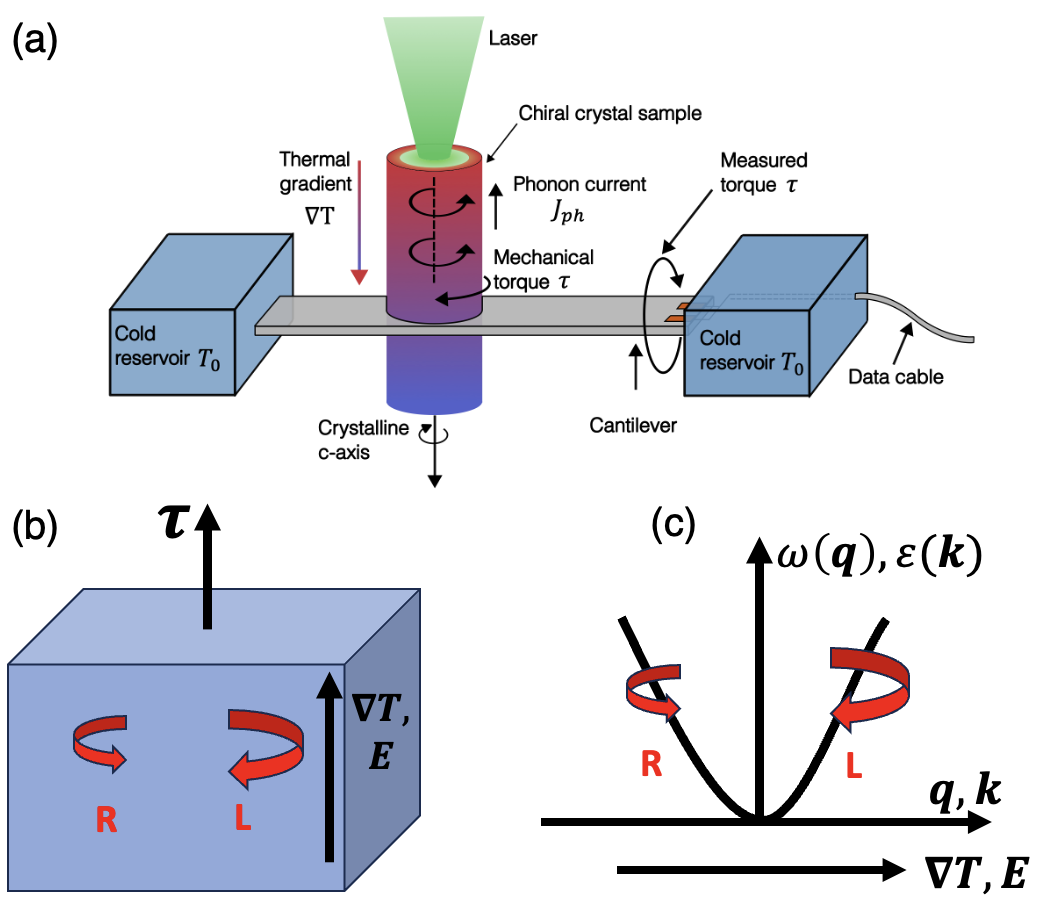}
    \caption{ (a) Schematic of the principal 'chiralometer' setup. Uncompensated angular momentum is excited by external perturbations. (b) A chiral phononic (or electronic) system with broken inversion symmetry hosts a compensated total angular momentum at equilibrium due to time-reversal symmetry. When driven out of equilibrium, the system acquires an uncompensated angular momentum, resulting in a measurable mechanical torque. (c) Non-equilibrium population imbalance between opposite chiralities.}
    \label{fig:setup}
\end{figure}

By applying external perturbations such as a temperature gradient for insulators or an electric field for metals, we generate uncompensated angular momentum within the corresponding excitations. To satisfy the conservation of total angular momentum, this internal imbalance must induce a physical rotation of the crystal lattice, resulting in a measurable macroscopic mechanical torque.

We calculate this induced torque for both insulating (phonon-mediated) and conducting (electron-mediated) systems. Specifically, we provide quantitative predictions for a series of chiral insulators, including Te, SiO$_2$, RhSi, PtAl, PdGa, PtGa, PdAl, as well as the topological semimetal CoSi. Our findings establish mechanical torque as a robust and experimentally feasible order parameter for chirality. When integrated with traditional spin-transport techniques, this mechanical probe offers a transformative tool for investigating a wide range of physical phenomena, particularly those central to the emerging field of orbitronics~\cite{orbitronics_review}.


\textbf{Chirality probe for insulators.} We begin by considering the chirality probe for insulators. In this case, due to the absence of electronic degrees of freedom, the main carriers of angular momentum are chiral phonons. Because phonons carry no electric charge, they are most effectively excited by a temperature gradient (which can be conveniently induced by shining a laser beam; see \cite{exper}). Other means of phonon excitation (including dragging of phonons by a limited number of conducting electrons) are also discussed below.

\begin{figure*}
    \centering
    \includegraphics[width=\linewidth]{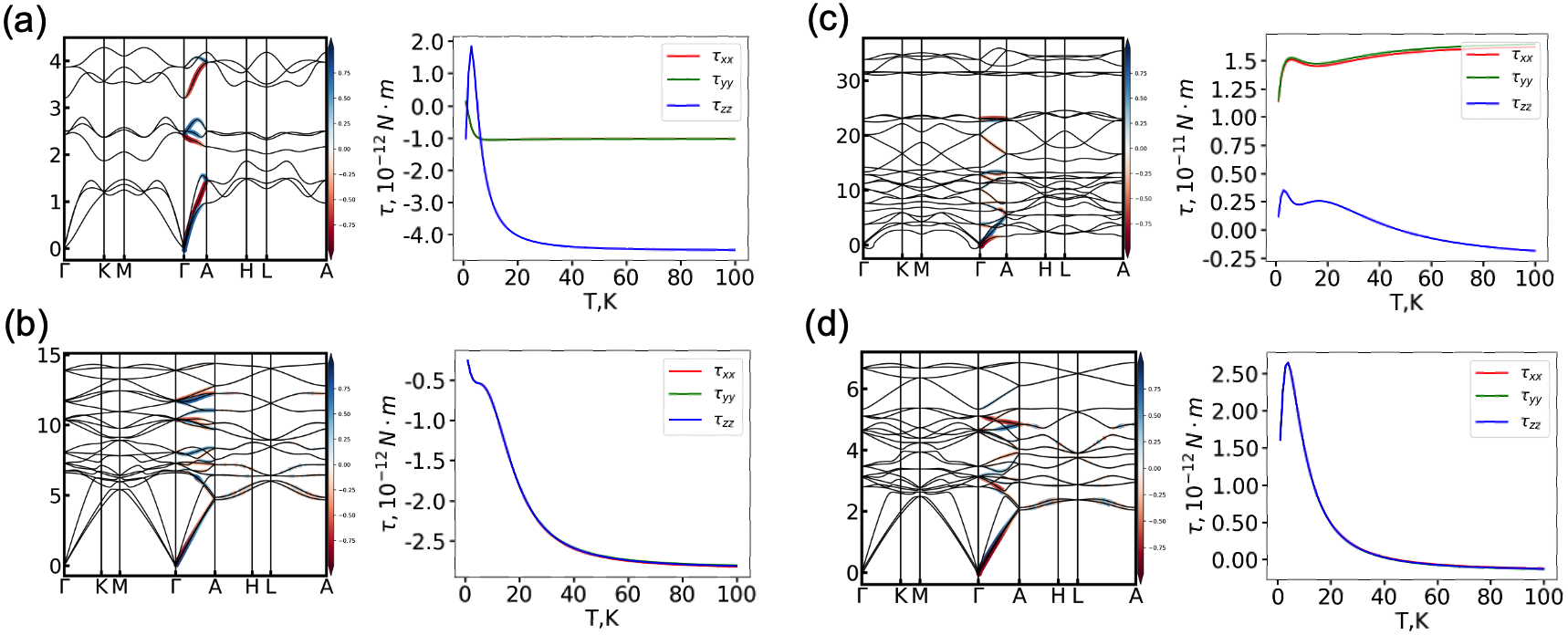}
    \caption{Calculated phonon dispersion and mechanical torque components for chiral crystals of (a) Te, (b) CoSi, (c) SiO$_2$, (d) PdGa. We assume a temperature gradient of $\nabla T=100\,\mathrm{K}/\mathrm{m}$ applied along the $x$, $y$ or $z$ axis (indicated by different colors) for a sample with dimensions $500\,\mu\mathrm{m}\times200\,\mu\mathrm{m}\times100\,\mu\mathrm{m}$. Here, $\tau_{ii}$ denotes the $i$-th torque component when $\nabla T$ is applied along the $x_i$ axis. Off-diagonal torque components are symmetry-forbidden \cite{phonon_ang_momentum}. The colorbar represents the angular momentum of each phonon branch.}
    \label{fig:phonon_results}
\end{figure*}

\textbf{Excitation with temperature gradient.}
A nonequilibrium phonon distribution function can be obtained from the Boltzmann equation as
\begin{align}
    \delta n_{\sigma,\mathbf{q}}=-\tau_\mathrm{rel} (\mathbf{v}_\sigma(\mathbf{q})\cdot\nabla T)\partial_Tn_\mathrm{eq}(\omega_{\sigma,\mathbf{q}}),\nonumber\\ n_\mathrm{eq}(\omega_{\sigma,\mathbf{q}})=(e^{\hbar\omega_{\sigma,\mathbf{q}}/k_BT}-1)^{-1}
    \label{eq:delTa_n},
\end{align}
where $\omega_{\sigma,\mathbf{q}}$ and $\mathbf{v}_\sigma(\mathbf{q})$ are the dispersion of phonon band $\sigma$ and the corresponding group velocity, respectively, and $\tau_\mathrm{rel}$ is the phonon relaxation time. The total angular momentum of phonons $\mathbf{L}^\mathrm{ph}$ is given by \cite{EdH_angular_phonon,PhysRevLett.132.056302,phonon_ang_momentum}:
\begin{align}
    \mathbf{L}^\mathrm{ph}=\sum_{\sigma}\int\frac{d^3\mathbf{q}}{(2\pi)^3}\mathbf{l}_{\sigma}(\mathbf{q})\left[n(\omega_{\sigma,\mathbf{q}})+\frac{1}{2}\right],\nonumber\\\mathbf{l}_\sigma(\mathbf{q})=i\hbar\mathbf{e}_\sigma(\mathbf{q})\times\mathbf{e}_\sigma^{*}(\mathbf{q}),
    \label{eq:ang_momentum}
\end{align}
with $\mathbf{l}_\sigma(\mathbf{q})$ being the angular momentum of band $\sigma$ at momentum $\mathbf{q}$ and $\mathbf{e}_\sigma(\mathbf{q})$ is the phonon mode polarization vector. Plugging the result of Eq. \eqref{eq:delTa_n} to Eq. \eqref{eq:ang_momentum} gives the following mechanical torque ${\bm\tau}=\delta \mathbf{L}^\mathrm{ph}/\tau_\mathrm{rel}$ (see also \cite{PhysRevLett.132.056302,phonon_ang_momentum}):
\begin{align}
    {\bm\tau}=-\sum_\sigma\int\frac{d^3\mathbf{q}}{(2\pi)^3}\mathbf{l}_{\sigma}(\mathbf{q})\left(\frac{\partial\omega_{\sigma,\mathbf{q}}}{\partial\mathbf{q}}\cdot\nabla T\right)\partial_T n_\mathrm{eq}(\omega_{\sigma,\mathbf{q}}).
    \label{eq:angular_momentum_result}
\end{align}
Based on results of Eq. \eqref{eq:angular_momentum_result}, we have performed real material calculations for chiral crystals Te, SiO$_2$, CoSi, RhSi, PtAl, PdGa, PtGa, PdAl (see Fig. \ref{fig:phonon_results}). The resulting mechanical torque for sample dimensions $500\,\mu\mathrm{m}\times200\,\mu\mathrm{m}\times100\,\mu\mathrm{m}$ is of the order of 
\begin{align}
    \tau\sim 10^{-11}\,\mathrm{N\cdot m},
\end{align}
which is well within experimentally feasible limits of $10^{-18}\,\mathrm{N}\cdot\mathrm{m}$ \cite{precision_limit}.

From Eqs. \eqref{eq:delTa_n}, \eqref{eq:angular_momentum_result} one readily sees that in the high-temperature limit $k_BT\gg\hbar\omega_D$ the mechanical torque becomes $T$-independent. In this regime, the equilibrium distribution function satisfies $n_\mathrm{eq}(\omega)\approx k_BT/\hbar\omega$ and its temperature derivative $\partial_T n_\mathrm{eq}(\omega)$ is therefore temperature independent. In the opposite limit of low temperatures, the torque vanishes due to the absence of thermally excited phonons in equilibrium. Within the intermediate temperature range, the torque behavior depends on angular momentum distribution of the relevant acoustic phonon bands. For some materials, this may result in a non-monotonic behavior with temperature. In the case of SiO$_2$, for instance, the acoustic bands separate into distinct contributions of positive and negative angular momentum (see Fig. \ref{fig:phonon_results} c). Consequently, the mechanical torque temperature dependence changes when the thermal energy exceeds the maximum energy of only one band, but not the other.

\textbf{Drag of phonons by electrons\label{phonon_electron_drag}}
Despite the absence of charge degrees of freedom, phonon modes can be nevertheless excited by an electric field. This excitation process is mediated by first driving the electrons (present in small but finite numbers even in an insulator) out of equilibrium. These non-equilibrium electrons would in turn drag the phonons, thereby generating a finite phonon mechanical torque:
\begin{align}
    {\bm\tau}=\sum_{\sigma,\mathbf{q}}\mathbf{l}_\sigma(\mathbf{q})\frac{\delta n_{\sigma,\mathbf{q}}}{\tau_\mathrm{e}}, \quad \delta n_{\sigma,\mathbf{q}}\approx- \mathbf{q\cdot u_\mathbf{q}}\,\frac{\partial n_{\sigma,\mathbf{q}}}{\partial\omega_{\sigma,\mathbf{q}}},
\end{align}
where $\tau_\mathrm{e}$ is electron's relaxation time and the phonon drift velocity $\mathbf{u_q}$ is defined by the product of electron-phonon coupling constant and driving electric field, detailed in the appendix \ref{app:phonon_drift}. The corresponding mechanical torque contribution then reads
\begin{align}
    {\bm\tau}=-\sum_\sigma\int\frac{d^3\mathbf{q}}{(2\pi)^3}\mathbf{l}_\sigma(\mathbf{q})(\mathbf{q\cdot u_q})\tau_\mathrm{e}^{-1}\partial_{\omega} n_\mathrm{eq}(\omega_{\sigma,\mathbf{q}}).
\end{align}
However, due to the low electronic density of electronic states in an insulator, the  electron-phonon coupling constant and the resulting drift velocity are strongly suppressed, rendering this mechanism negligible compared to the purely phonon contribution. This contribution could nevertheless be made observable in a material with strong spin-orbit coupling subjected to a high external electric field.

\textbf{Chirality measurement for metals.}
The idea of probing chirality by driving angular-momentum carriers out of equilibrium can be extended to conducting systems as well. However, the high thermal conductivity of such systems makes it difficult to establish a sufficiently large temperature gradient. To circumvent this challenge and isolate the electronic contribution, this section focuses on electric-field driving. While the electron--phonon coupling discussed in the previous section ``Drag of phonons by electrons'' (and a distinct excitation channel in which electron spins couple to phonon angular momentum~\cite{2025phonon}) can also excite chiral phonon modes, the present analysis is confined exclusively to contributions from electronic degrees of freedom.

In what follows, we model electronic chirality in two ways: either as a system with structural chirality (a helical molecule) or as a system with a topological band crossing and nontrivial Berry curvature that induces orbital angular momentum. Similar phenomena have previously been considered in both the structural- and topological-chirality settings. Namely, sensing chirality of a helical structure by driving a magnetic moment was studied in \cite{current_induced_magnetization}, and the gyromagnetic effect in a Weyl semimetal was discussed in \cite{analytic_ang_momentum}.


\textbf{Structural chirality}
We first demonstrate that structural chirality can induce a non-trivial mechanical torque. For that, we employ a simple tight-binding model of a quasi-1D chiral crystal. This model consists of three atoms per unit cell arranged along a helical line, which inherently breaks inversion symmetry. We adopt the specific formulation from \cite{helical_molecule}, in which each atom hosts three $l=1$ atomic orbitals and only nearest-neighbor hopping between atoms is allowed. A detailed definition of the model Hamiltonian is provided in appendix \ref{app:helical_molecule}. The induced mechanical torque per unit length is then oriented along the $z$ axis and is given by
\begin{align}
    \tau_z = \sum_n\int\frac{dk}{2\pi}e(\mathbf{v}_n(k)\mathbf{E})\left[-\partial_\varepsilon f_\mathrm{eq}(E_n\{k\})\right]l^n_z(k),\nonumber\\
    l^n_z(k)=\langle\psi_n(k)|\hat{L}_z|\psi_n(k)\rangle,
\end{align}
where $\hat{L}_z$, $l_z^n(k)$ are the $z$ component $L=1$ angular momentum operator and its expectation value for band $n$ and $f_\mathrm{eq}(E_n(k))$ is Fermi distribution function.

\begin{figure}
    \centering
    \includegraphics[width=\linewidth]{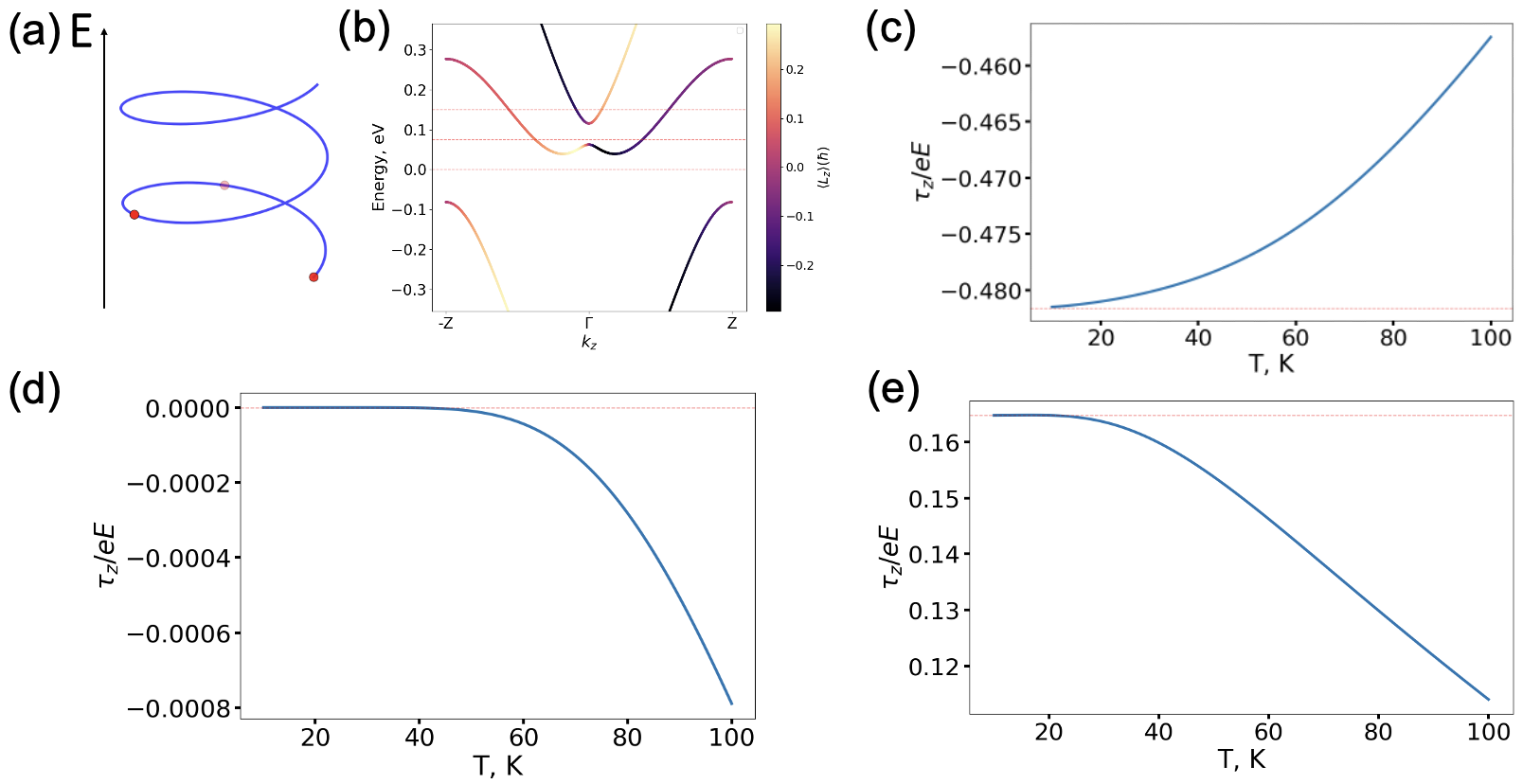}
    \caption{(a) Schematics of a helical molecule with three atoms per unit cell. As a minimal non-trivial model, each site hosts three orbitals with angular momentum $l=1$.
    (b) Band structure for varying chemical potential positions, with the angular momentum of each band indicated by color. (c) - (e) Calculated mechanical torque for chemical potential values of $\mu =0.15\,\mathrm{eV}$, $0\,\mathrm{eV}$ and $0.075\,\mathrm{eV}$, respectively. The zero temperature saturation values are indicated by red dashed lines.}
    \label{fig:helical_molecule}
\end{figure}

The computational results for different positions of chemical potential are presented in Fig. \ref{fig:helical_molecule}. 
They indicate either exponential (for the chemical potential lying within the band gap) or power law scaling with temperature $T$. Unlike the phonon case, the total electron number is temperature independent. Consequently, the electronic contribution does not vanish at the lowest temperatures. Furthermore, we expect these results to be robust against weak disorder, since the momentum relaxation time cancels out from the final expression. The proposed configuration could be realized for molecular crystals \cite{molecular_crystal1,molecular_crystal2,molecular_crystal3}. Besides molecular crystals, an analogous phenomenon (known as the Lehmann effect) is already known for liquid crystals \cite{lehmann1,lehmann2}.

\textbf{Topological chirality}
In the context of condensed matter physics, the notion of electronic chirality typically refers to topological band crossing, non-trivial Berry curvature and Chern number. In this case one can introduce electric field-induced uncompensated angular momentum of electrons by
\begin{align}
    \mathbf{J}_\mathrm{e}=\sum_n\int\frac{d^3\mathbf{k}}{(2\pi)^3}\mathbf{l}_n(\mathbf{k})\,\delta f_n(\mathbf{k})=\nonumber\\=\sum_n\int\frac{d^3\mathbf{k}}{(2\pi)^3}\mathbf{l}_n(\mathbf{k})\,e\left(\mathbf{E}\mathbf{v}_n\right)\tau\left(-\partial_\varepsilon f_\mathrm{eq}(E_n(\mathbf{k}))\right),
    \label{eq:torque_topology}
\end{align}
where $n$, $\mathbf{k}$ denote electronic bands and momentum correspondingly,  $\mathbf{v}_n=\partial_{\hbar\mathbf{k}}E_n(\mathbf{k})$ and orbital magnetic moment $\mathbf{l}_n(\mathbf{k})$ is defined by \cite{orb_momentum_berry_phase}
\begin{align}
    \mathbf{l}_n(\mathbf{k}) = -i\frac{e}{2\hbar\gamma}\langle \nabla_\mathbf{k} \psi_n|\times \left(H(\mathbf{k})-E_n(\mathbf{k})\right)|\nabla_\mathbf{k}\psi_n\rangle,
\end{align}
where $\gamma$ is gyromagnetic ratio. For simplicity we assume the isotropic Weyl Hamiltonian with quadratic terms
\begin{align}
    H(\mathbf{k})=\frac{\hbar^2\mathbf{k}^2}{2m^*}\sigma_0+\hbar v_F\mathbf{k}\cdot{\bm\sigma},
    \label{eq:ham_isotropic}
\end{align}
so that the gyromagnetic ratio $\gamma$ for this Hamiltonian is $\gamma=\frac{ge\hbar}{2m^*}$. The resulting mechanical torque per unit volume can then be expressed as
\begin{align}
    {\bm \tau} = \frac{e^2}{12\pi^2\hbar^2\gamma}C\mathbf{E}(\mu_+-\mu_-),
    \label{eq:torque_electrons_analytical}
\end{align}
where $C=\pm1$ denotes the Chern number of the node. Similar to the structural chirality case, the result in Eq. \eqref{eq:torque_electrons_analytical} requires only the breaking of inversion symmetry, which ensures $\mu_+\ne\mu_-$ for opposite chirality nodes. Therefore, we anticipate that this mechanical torque probe of topology should be applicable to both magnetic and non-magnetic noncentrosymmetric Weyl semimetals. To illustrate this conclusion we provide two complementary demonstrations: a realistic material calculation for CoSi and a two band lattice model calculation. A complete analysis of the symmetry properties governing  the torque response (valid both for conducting and insulating systems) could be found in \cite{phonon_ang_momentum,exper}. 

\textbf{Two-band lattice model and first-principles calculation of CoSi.}
We define a minimal two-band lattice model featuring topological Weyl crossings as follows:
\begin{align}
    H=d_0(\mathbf{k})\sigma_0+\mathbf{d}(\mathbf{k})\cdot{\bm\sigma},\nonumber\\
    d_0(\mathbf{k})=\gamma\sin k_3,\quad \mathbf{d}(\mathbf{k})=\begin{pmatrix}
        t\sin k_1\\
        t\sin k_2\\
        M-t\sum_{i=1}^3 \cos k_i
    \end{pmatrix}.
\end{align}
Two topological Weyl points occur for $1<M<3$ located at $k_x=k_y=0$ and $k_z=\arccos(M/t-2),\,\pi-\arccos(M/t-2)$. Although this model explicitly breaks both time reversal and inversion symmetry, a four-band time reversal invariant lattice model can be constructed by 'doubling' the original model with its time reversal counterpart. The resulting non-equilibrium mechanical torque, defined by Eq. \eqref{eq:torque_topology}, is even under time reversal and would thus be also doubled. 

\begin{figure}
    \centering
    \includegraphics[width=\linewidth]{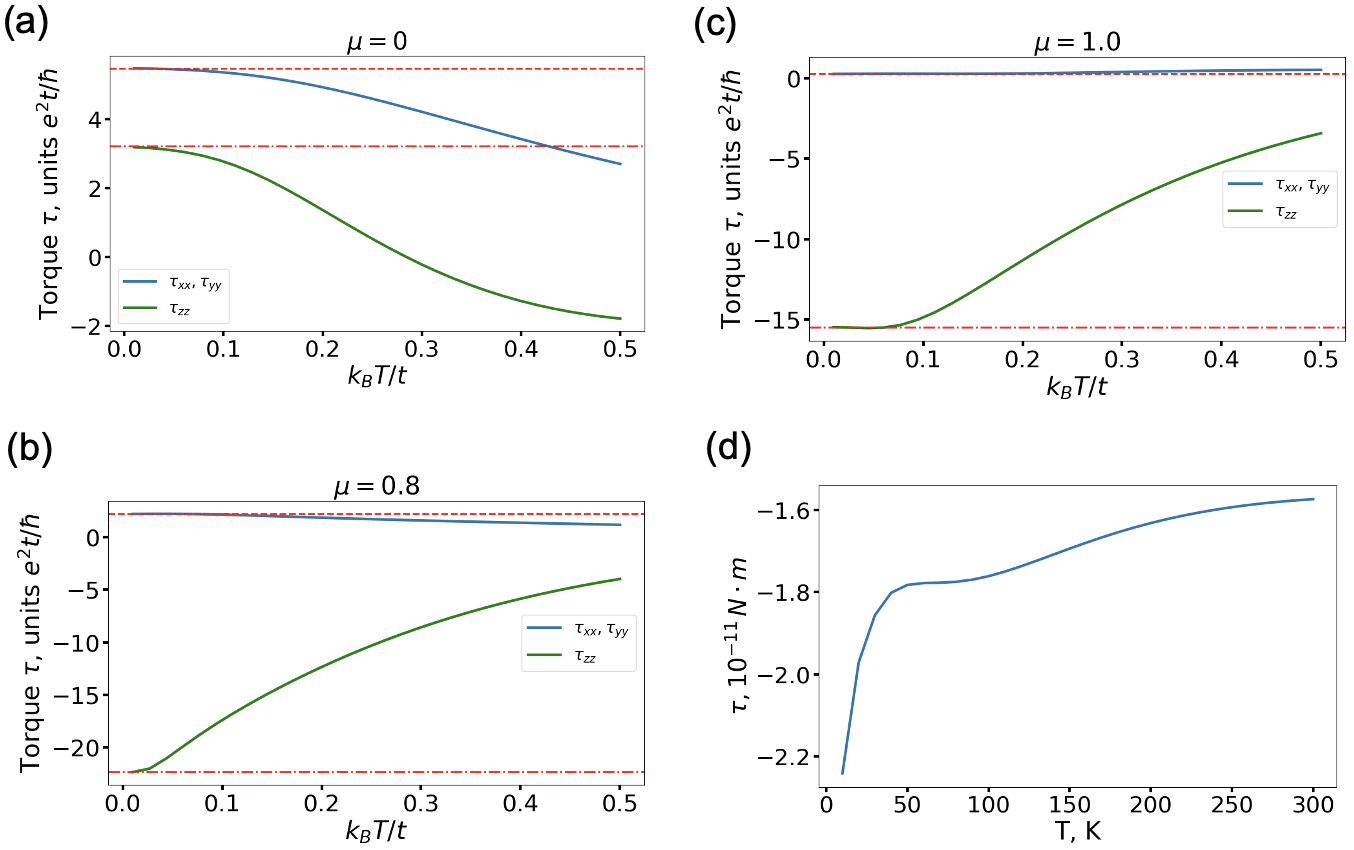}
    \caption{Results of mechanical torque calculations for a two-band model and the real material CoSi. Panels (a)-(c) display the two-band model results with the chemical potential $\mu$ set to (a) $\mu=0$,(b) $\mu=0.8$, (c) $\mu=1.0$; the zero temperature saturation values are indicated by red dashed lines. Panel (d) presents the real material results for CoSi, corresponding to the angular momentum induced by electronic orbital motion. The calculations assume a driving electric field $E=10^4\,\mathrm{V/m}$ and crystal dimensions $1\,\mathrm{mm}\times 0.2\,\mathrm{mm}\times0.1\,\mathrm{mm}$. Owing to the cubic symmetry of CoSi, the response is isotropic, i.e., $\tau_{xx}=\tau_{yy}=\tau_{zz}$.}
    \label{fig:two_band}
\end{figure}

In the calculations presented below, we have used the parameter values $M=2$ and $\gamma=0.8$. The results for all torque components (per unit volume) for different chemical potential positions are shown in Fig. \ref{fig:two_band}. 
In accord with the structural chirality discussion, for all parameter values predicted mechanical torque saturates at low temperatures.

We further performed a real material calculation for the chiral crystal CoSi, illustrated in Fig. \ref{fig:two_band}d. Assuming a CoSi crystal of reasonable dimensions $1\,\mathrm{mm}\times 0.2\,\mathrm{mm}\times0.1\,\mathrm{mm}$, a $g$ factor of $10$ (large $g$ factors in the range of $5-40$ have been reported for similar topological semimetals \cite{cd3as2_g,tap_g,zrsis,zrte5_g}) and a driving electric field $E=10^4\,\mathrm{V/m}$, the resulting torque is of the order of $10^{-11}\,\mathrm{N\cdot m}$. This value is comparable to phonon-induced torque and lies well within experimental detectability.

\textbf{Optical excitation}
As a side remark, we note that an applied electric field can also induce vertical interband optical transition of electrons. The resulting mechanical torque from such transitions is proportional to both the vertical transition rate $\Gamma$ and the imbalance between left- and right-handed electrons in the material. Under an external field $E$, the electron chirality imbalance is linear in $E$, while the vertical transition rate scales quadratically as $\Gamma\propto E^2$, leading to a torque $\tau\propto E^3$. This strong cubic dependence renders the effect negligible compared to the linear-in-$E$ contribution discussed earlier. We further note that a related phenomenon, namely mechanical torques excited by circularly or linearly polarized light, has been previously studied in \cite{optical_driving1,optical_driving2,theory_torque}.



\textbf{Conclusions.}
In this study, we developed a direct mechanical method for probing chirality in both insulating and conducting crystalline systems. By driving the system out of equilibrium (either through a temperature gradient in insulators or an electric field in conductors), we generate a measurable mechanical torque. This torque arises from the uncompensated angular momentum of chiral phonons or electrons, offering a clear and robust experimental signature of structural and topological chirality.

Our calculations suggest that the predicted torque magnitudes are well within experimentally accessible ranges ($10^{-11}\,\mathrm{N\cdot m}$), supporting the feasibility of this approach. Notably, the proposal for insulating crystals has already been validated in recent experiments on Te~\cite{exper}, confirming the viability of phonon angular-momentum measurements. For conducting systems, we extended this framework to electronic chirality, encompassing both structurally chiral systems (such as helical molecules and molecular crystals) and topologically chiral materials (e.g., the topological semimetal CoSi).

This work establishes mechanical torque as a new order parameter for chirality that complements existing optical and spin-based techniques. By enabling direct chirality detection across a wide range of materials, our approach paves the way for deeper exploration of chiral phenomena in condensed matter, with potential applications in orbitronics, spintronics, and the study of topological phases.

\textbf{Acknowledgments.}
N.M. acknowledges financial support from the Alexander von Humboldt Foundation. N.M., N.P., and C.F. acknowledge financial support from the Deutsche Forschungsgemeinschaft (DFG, German Research Foundation) through the Würzburg--Dresden Cluster of Excellence ctd.qmat -- Complexity, Topology and Dynamics in Quantum Matter (EXC 2147, project-id 390858490).

\nocite{*}

\bibliography{biblio}
\clearpage

\appendix
\newpage 

\section{Drag of phonons by electrons generating mechanical torque \label{app:phonon_drift}}
In this appendix, we demonstrate how mechanical torque can be generated by dragging chiral phonons with nonequilibrium electrons via electron--phonon coupling. Namely, due to the extremely low number of electrons in an insulator, electronic excitations can be more long-lived than phonons: $\tau_\mathrm{e}\gg\tau_\mathrm{ph}$. This, in turn, makes it possible to excite phonon degrees of freedom by first exciting electrons with an external electric field and then coupling this excitation to phonons, driving them out of equilibrium and producing a mechanical torque.

We do this by solving coupled semiclassical Boltzmann equations for the phononic $n(\mathbf{q})$ and electronic $f(\mathbf{k})$ distribution functions. Our end result for $n(\mathbf{q})$ contains a ``Doppler shift'' drag term:
\begin{align}
    n(\mathbf{q})=n_\mathrm{eq}(\omega(\mathbf{q})-\mathbf{q}\cdot\mathbf{u}_\mathbf{q}),
    \label{eq:drift_phonons}
\end{align}
with $n_\mathrm{eq}(\varepsilon)$ the Bose distribution function. The corresponding Boltzmann equations then read
\begin{align}
    e\mathbf{E}\cdot\mathbf{v}(\mathbf{k})\partial_\varepsilon f_\mathrm{eq}=I_\mathrm{e-imp}+I_\mathrm{e-ph},\quad I_\mathrm{e-imp}=-\frac{f-f_\mathrm{eq}}{\tau_\mathrm{e}},\nonumber\\
    I_\mathrm{ph-e}+I_\mathrm{ph-ph}+I_\mathrm{ph-imp}=0.
\end{align}
\begin{widetext}
Electron-phonon scattering contribution to phonon collision integral is given by
\begin{align}
    I_\mathrm{e-ph}\{n_\mathbf{q}\}=\sum_\mathbf{k}\left[W^\mathrm{in}_\mathbf{k+q,k}-W^\mathrm{out}_\mathbf{k+q,k}\right],\nonumber\\
    W^\mathrm{in}_\mathbf{k+q,k}=\frac{2\pi}{\hbar}|g_\mathbf{q}|^2f_\mathbf{k+q}(1-f_\mathbf{k})(1+n_\mathbf{q})\delta(\varepsilon_\mathbf{k+q}-\varepsilon_\mathbf{k}-\hbar\omega_\mathbf{q}),\nonumber\\ W^\mathrm{out}_\mathbf{k,k+q}=\frac{2\pi}{\hbar}|g_\mathbf{q}|^2f_\mathbf{k}(1-f_\mathbf{k+q})n_\mathbf{q}\delta(\varepsilon_\mathbf{k+q}-\varepsilon_\mathbf{k}-\hbar\omega_\mathbf{q}).
\end{align}
In what follows we derive phonon-electron collision integral partly based on a classical textbook derivation \cite{Landau_physical_kinetics}. As a next step we assume equilibrium distribution function for phonons and prove that electrons, while being driven by electric field, can drag phonons along, resulting in non-equilibrium shift of phonon distribution function and final mechanical torque due to phonons. For the non-equilibrium electrons distribution function we have
\begin{align}
    \delta f_\mathbf{k}=-e(\mathbf{v}(\mathbf{k})\cdot\mathbf{E})\tau_\mathrm{e}\partial_{\varepsilon_\mathbf{k}} f_\mathrm{eq}(\varepsilon_\mathbf{k})\equiv(-\partial_{\varepsilon_\mathbf{k}}f_\mathrm{eq}(\varepsilon_\mathbf{k}))\Phi_\mathbf{k}=\frac{1}{k_BT}f_\mathbf{eq}(\varepsilon_\mathbf{k})(1-f_\mathbf{eq}(\varepsilon_\mathbf{k}))\Phi_\mathbf{k},\quad \mathbf{v}(\mathbf{k})=\partial_{\hbar\mathbf{k}}\varepsilon(\mathbf{k}),
\end{align}
where $\tau_\mathrm{e}$ stands for the electron momentum relaxation time (e.g., due to impurity scattering) and $\varepsilon(\mathbf{k})=\hbar^2\mathbf{k}^2/2m^*$ is, for simplicity, taken to describe a spherically symmetric Fermi pocket. Transforming the electron--phonon collision integral, one arrives at
\begin{align}
    I_\mathrm{e-ph}\{n_\mathbf{q}\}=\frac{2\pi}{\hbar}|g_\mathbf{q}|^2\sum_\mathbf{k}\big[f_\mathbf{k+q}(1-f_\mathbf{k})(n_\mathbf{q}+1)-f_\mathbf{k}(1-f_\mathbf{k+q})n_\mathbf{q}\big]\delta(\varepsilon_\mathbf{k+q}-\varepsilon_\mathbf{k}-\hbar\omega_\mathbf{q}).
\end{align}
For equilibrium distribution functions $n_\mathrm{eq}$ and $f_\mathrm{eq}$, the collision integral $I_\mathrm{e-ph}\{n_\mathbf{q}\}$ vanishes due to the relation
\begin{align}
    \frac{1+n_\mathrm{eq}(\varepsilon'-\varepsilon)}{n_\mathrm{eq}(\varepsilon'-\varepsilon)}=\frac{f_\mathrm{eq}(\varepsilon)}{1-f_\mathrm{eq}(\varepsilon)}\frac{1-f_\mathrm{eq}(\varepsilon')}{f_\mathrm{eq}(\varepsilon')}.
\end{align}
After expanding $f=f_\mathrm{eq}+\delta f$ one has
\begin{align}
    I_\mathrm{e-ph}\{n_\mathbf{q}\}=\frac{2\pi}{\hbar}|g_\mathbf{q}|^2\sum_\mathbf{k}(1+n_\mathbf{q})(1-f_\mathbf{k})(1-f_\mathbf{k+q})\left[\frac{f_\mathbf{k+q}}{1-f_\mathbf{k+q}}-\frac{n_\mathbf{q}}{1+n_\mathbf{q}}\frac{f_\mathbf{k}}{1-f_\mathbf{k}}\right]\delta(\varepsilon_\mathbf{k+q}-\varepsilon_\mathbf{k}-\hbar\omega_\mathbf{q})\approx\nonumber\\\approx\frac{2\pi}{\hbar}|g_\mathbf{q}|^2\sum_\mathbf{k}(1+n_\mathbf{q})(1-f_\mathbf{k})(1-f_\mathbf{k+q})\left[\frac{\delta f_\mathbf{k+q}}{(1-f_\mathbf{k+q})^2}-\frac{n_\mathbf{q}}{1+n_\mathbf{q}}\frac{\delta f_\mathbf{k}}{(1-f_\mathbf{k})^2}\right]\delta(\varepsilon_\mathbf{k+q}-\varepsilon_\mathbf{k}-\hbar\omega_\mathbf{q})=\nonumber\\=\frac{2\pi}{\hbar}|g_\mathbf{q}|^2\sum_\mathbf{k}\frac{(1+n_\mathbf{q})(1-f_\mathbf{k})(1-f_\mathbf{k+q})}{k_BT}\left[\frac{f_\mathbf{k+q}}{1-f_\mathbf{k+q}}\Phi_\mathbf{k+q}-\frac{n_\mathbf{q}}{1+n_\mathbf{q}}\frac{f_\mathbf{k}}{1-f_\mathbf{k}}\Phi_\mathbf{k}\right]\delta(\varepsilon_\mathbf{k+q}-\varepsilon_\mathbf{k}-\hbar\omega_\mathbf{q})=\nonumber\\=\frac{2\pi}{\hbar}|g_\mathbf{q}|^2\sum_\mathbf{k}\frac{(1+n_\mathbf{q})(1-f_\mathbf{k})f_\mathbf{k+q}}{k_BT}\left[\Phi_\mathbf{k+q}-\Phi_\mathbf{k}\right]\delta(\varepsilon_\mathbf{k+q}-\varepsilon_\mathbf{k}-\hbar\omega_\mathbf{q}).
\end{align}
Employing the identity
\begin{align}
    n_\mathrm{eq}(\varepsilon'-\varepsilon)=-\frac{f_\mathrm{eq}(\varepsilon')(1-f_\mathrm{eq}(\varepsilon))}{f_\mathrm{eq}(\varepsilon')-f_\mathrm{eq}(\varepsilon)},
\end{align}
the expression for $I_\mathrm{e-ph}\{n_\mathbf{q}\}$ could be rewritten as
\begin{align}
    I_\mathrm{e-ph}\{n_\mathbf{q}\}=-\frac{2\pi}{\hbar}|g_\mathbf{q}|^2\sum_\mathbf{k}\frac{n_\mathbf{q}(1+n_\mathbf{q})}{k_BT}(f_\mathbf{k+q}-f_\mathbf{k})[\Phi_\mathbf{k+q}-\Phi_\mathbf{k}]\delta(\varepsilon_\mathbf{k+q}-\varepsilon_\mathbf{k}-\hbar\omega_\mathbf{q}).
\end{align}
Since $q_D\ll k_F$, the integral over $\mathbf{k}$ comes from the vicinity of Fermi surface: $f(\varepsilon_\mathbf{k+q})-f(\varepsilon_\mathbf{k})\approx\partial_\varepsilon f_\mathrm{eq}(\varepsilon_\mathbf{k})\hbar\omega_\mathbf{q}$ and
\begin{align}
    I_\mathrm{e-ph}\{n_\mathbf{q}\}\approx\frac{2\pi}{\hbar}|g_\mathbf{q}|^2\frac{n_\mathbf{q}(1+n_\mathbf{q})}{k_BT}\hbar\omega_\mathbf{q}\int d\varepsilon_\mathbf{k}\,\nu(\varepsilon_\mathbf{k}) (-\partial_{\varepsilon_\mathbf{k}}f_\mathrm{eq}(\varepsilon_\mathbf{k}))e\tau_\mathrm{e}\frac{\hbar\mathbf{q}}{m*}\cdot\mathbf{E}\,\frac{m^*}{\hbar^2|\mathbf{k}||\mathbf{q}|}=\nonumber\\=\frac{2\pi}{\hbar}|g_\mathbf{q}|^2\frac{n_\mathbf{q}(1+n_\mathbf{q})}{k_BT}\hbar\omega_\mathbf{q}\nu_F\frac{e\tau_e\,\mathbf{q}\cdot\mathbf{E}}{\hbar k_F|\mathbf{q}|}.
\end{align}
Now we solve the equation for phonon distribution function
\begin{align}
    I_\mathrm{e-ph}\{ n_\mathbf{q}\}-\frac{\delta n_\mathbf{q}}{\tau_\mathrm{ph}}=0,
\end{align}
with $\tau_\mathrm{ph}$ being phonon relaxation time that accounts for all the relaxation processes except for electron-phonon scattering (meaning phonon-phonon and phonon-impurity scattering). For $\delta n_\mathbf{q}$ one finally arrives at
\begin{align}
    \delta n_\mathbf{q}=(-\partial_{\omega_\mathbf{q}} n_\mathrm{eq}(\omega_\mathbf{q}))\mathbf{q\cdot u}_\mathbf{q},\quad \mathbf{u}_\mathbf{q}=\frac{2\pi}{\hbar^2}|g_\mathbf{q}|^2\hbar\omega_\mathbf{q}\nu_F\tau_\mathrm{ph}\frac{e\mathbf{E}\tau_\mathrm{e}}{p_F}\frac{1}{|\mathbf{q}|}.
\end{align}
We now estimate by the order of magnitude the resulting mechanical torque from phonon drag by electrons. The expression for mechanical torque ${\bm\tau}$ reads
\begin{align}
    {\bm\tau}=V\sum_\sigma\int\frac{d^3\mathbf{q}}{(2\pi)^3}\mathbf{l}_\sigma(\mathbf{q})\frac{2\pi}{\hbar^2}\lambda_{\sigma,\mathbf{q}}\hbar^2\omega^2_{\sigma,\mathbf{q}}\frac{\tau_\mathrm{ph}}{\tau_\mathrm{e}}\frac{e\mathbf{q}\cdot\mathbf{E}\tau_\mathrm{e}}{p_F|\mathbf{q}|}(-\partial_{\omega_{\sigma,\mathbf{q}}} n_\mathrm{eq}(\omega_{\sigma,\mathbf{q}})),\quad \lambda_{\sigma,\mathbf{q}}=\frac{|g_{\sigma,\mathbf{q}}|^2\nu_F}{\hbar\omega_{\sigma,\mathbf{q}}}
\end{align}
so its order of magnitude estimate is given by (we expect that the main contribution comes from acoustic phonons with dispersion $\omega_\mathbf{q}=\hbar s|\mathbf{q}|$)
\begin{align}
    \tau\sim V\nu_\mathrm{ph}(k_BT)k_BT\hbar\frac{1}{\hbar^2}\lambda(k_BT)^2\frac{\tau_\mathrm{ph}}{\tau_\mathrm{e}}\frac{eE\tau_e}{p_F}\frac{\hbar}{k_BT}\sim V\cdot \frac{(k_BT)^4}{(\hbar s)^3}\frac{eE\tau_e}{p_F}\frac{\tau_\mathrm{ph}}{\tau_\mathrm{e}}\lambda.
\end{align}
For sample dimensions $1\,\mathrm{mm}\times0.2\,\mathrm{mm}\times0.1\,\mathrm{mm}$, $T=10$ K and the speed of sound $s\sim10^3\,\mathrm{m/s}$ the resulting mechanical torque is
\begin{align}
    \tau\sim10^{-6}\frac{eE\tau_e}{p_F}\frac{\tau_\mathrm{ph}}{\tau_\mathrm{e}}\lambda\,[\mathrm{N\cdot m}].
\end{align}
Given that the applicability criteria for phonon drag effect require both $eE\tau_e/p_F\ll1$, $\tau_\mathrm{ph}/\tau_\mathrm{e}\ll1$ weak electron-phonon coupling in insulators ($\lambda\ll1$ due to vanishing $\nu_F$), the described effect should be small compared to purely phonon contribution. Nonetheless, this effect could be made observable in materials with very strong spin-orbit coupling subject to high electric fields.
\end{widetext}

\section{Helical molecule model: tight-binding hamiltonian \label{app:helical_molecule}}
To describe the torque induced in a helical molecule we adopt a tight-binding model from \cite{helical_molecule}. Its Hamiltonian reads:
\begin{align}
    H(k)=\begin{pmatrix}
        0 & T_{12}e^{-ika/3} & T_{13}e^{ika/3}\\
        T_{21}e^{ika/3} & 0 & T_{23}e^{-ika/3}\\
        T_{31}e^{-ika/3} & T_{32}e^{ika/3} & 0
    \end{pmatrix},
\end{align}
with nearest-neighbor hopping matrix $T_{ij}$ given by
\begin{align}
    T_{ij}(\phi_{ij},\theta_{ij})=\begin{pmatrix}
        t_{ij,xx} & t_{ij,xy} & t_{ij,xz}\\
        t_{ij,yx} & t_{ij,yy} & t_{ij,yz}\\
        t_{ij,zx} & t_{ij,zy} & t_{ij,zz}
    \end{pmatrix}.
\end{align}
We set hopping parameters to be $t_\sigma = 1.5\,\mathrm{eV}$ and $t_\pi = -0.5\,\mathrm{eV}$ correspondingly. We set offsite energies of the orbitals to be $\Delta_{p_x}=0.6\,\mathrm{eV}$, $\Delta_{p_y}=-0.4\,\mathrm{eV}$, $\Delta_{p_z}=0$ in order to avoid band crossing, leaving the discussion of band crossing effects to the study of the two-band model presented in the main text. The matrix elements $t_{ij}$ are given by
\begin{align}
    t_{ij,xx} = t_\pi\sin^2\phi_{ij}+\cos^2\phi_{ij} (t_\sigma \sin^2\theta_{ij}+t_\pi \cos^2\theta_{ij})\nonumber\\
    t_{ij,yy} = t_\pi\cos^2\phi_{ij}+\sin^2\phi_{ij} (t_\sigma \sin^2\theta_{ij}+t_\pi \cos^2\theta_{ij})\nonumber\\
    t_{ij,zz} = t_\sigma \cos^2\theta_{ij} + t_\pi\sin^2\theta_{ij}\nonumber\\
    t_{ij,xy} = t_{ij,yx} = \sin\phi_{ij}\cos\phi_{ij}\left(t_\sigma\sin^2\theta_{ij}-t_\pi\cos^2\theta_{ij}\right)\nonumber\\
    t_{ij,xz} = t_{ij,zx} = \cos\phi_{ij}\sin\theta_{ij}\cos\theta_{ij}(t_\sigma-t_\pi)\nonumber\\
    t_{ij,yz} = t_{ij,zy} = \sin\phi_{ij}\sin\theta_{ij}\cos\theta_{ij}(t_\sigma-t_\pi)
\end{align}
with spherical angles $\phi_{ij},\theta_{ij}$ describing the relative position of site $j$ with respect to site $i$. We assume hopping amplitudes to be $t_\sigma=1.5\,\mathrm{eV},\,t_\pi=-0.5\,\mathrm{eV}$ and set $\theta_{ij}=\pi/4,3\pi/4$, $\phi_{ij}=\pm\pi/6,\pm5\pi/6,\pm3\pi/2$.

\section{DFT calculation}

Density functional theory (DFT) calculations are performed within the generalized gradient approximation (GGA), employing the Perdew-Burke-Ernzerhof (PBE) exchange-correlation functional~\cite{vasp}. A plane-wave energy cutoff of $500$ eV is adopted, and the Brillouin zone is sampled using an $8 \times 8 \times 8$ k-point mesh. Structural relaxations are carried out until the residual forces on each atom were less than $0.01$ eV/Å. Subsequently, force constants are calculated using density functional perturbation theory (DFPT). To ensure numerical convergence, the mechanical torques are evaluated on a dense $50 \times 50 \times 50$ k-point mesh.

\end{document}